\documentclass[twoside,twocolumn,9pt]{article}
\usepackage{extsizes}
\usepackage[super,sort&compress,comma]{natbib} 
\usepackage[version=3]{mhchem}
\usepackage[left=1.5cm, right=1.5cm, top=1.785cm, bottom=2.0cm]{geometry}
\usepackage{balance}
\usepackage{times,mathptmx}
\usepackage{sectsty}
\usepackage{graphicx} 
\usepackage{lastpage}
\usepackage[format=plain,justification=justified,singlelinecheck=false,font={stretch=1.125,small,sf},labelfont=bf,labelsep=space]{caption}
\usepackage{float}
\usepackage{fancyhdr}
\usepackage{fnpos}
\usepackage[english]{babel}
\addto{\captionsenglish}{%
  
}
\usepackage{array}
\usepackage{droidsans}
\usepackage{charter}
\usepackage[T1]{fontenc}
\usepackage[usenames,dvipsnames]{xcolor}
\usepackage{setspace}
\usepackage[compact]{titlesec}
\usepackage{hyperref}

\usepackage{siunitx}
\usepackage{amsmath}
\usepackage{amssymb}

\definecolor{cream}{RGB}{222,217,201}

\begin{document}

\pagestyle{fancy}
\thispagestyle{plain}
\fancypagestyle{plain}{

\renewcommand{\headrulewidth}{0pt}
}

\makeFNbottom
\makeatletter
\renewcommand\LARGE{\@setfontsize\LARGE{15pt}{17}}
\renewcommand\Large{\@setfontsize\Large{12pt}{14}}
\renewcommand\large{\@setfontsize\large{10pt}{12}}
\renewcommand\footnotesize{\@setfontsize\footnotesize{7pt}{10}}
\makeatother

\renewcommand{\thefootnote}{\fnsymbol{footnote}}
\renewcommand\footnoterule{\vspace*{1pt}%
\color{cream}\hrule width 3.5in height 0.4pt \color{black}\vspace*{5pt}} 
\setcounter{secnumdepth}{5}

\makeatletter 
\renewcommand\@biblabel[1]{#1}            
\renewcommand\@makefntext[1]%
{\noindent\makebox[0pt][r]{\@thefnmark\,}#1}
\makeatother 
\renewcommand{\figurename}{\small{Fig.}~}
\sectionfont{\sffamily\Large}
\subsectionfont{\normalsize}
\subsubsectionfont{\bf}
\setstretch{1.125} 
\setlength{\skip\footins}{0.8cm}
\setlength{\footnotesep}{0.25cm}
\setlength{\jot}{10pt}
\titlespacing*{\section}{0pt}{4pt}{4pt}
\titlespacing*{\subsection}{0pt}{15pt}{1pt}

\fancyfoot{}
\fancyfoot[LO,RE]{\vspace{-7.1pt}\includegraphics[height=9pt]{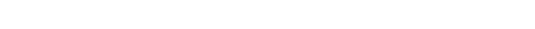}}
\fancyfoot[CO]{\vspace{-7.1pt}\hspace{13.2cm}\includegraphics{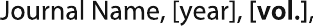}}
\fancyfoot[CE]{\vspace{-7.2pt}\hspace{-14.2cm}\includegraphics{head_foot/RF}}
\fancyfoot[RO]{\footnotesize{\sffamily{1--\pageref{LastPage} ~\textbar  \hspace{2pt}\thepage}}}
\fancyfoot[LE]{\footnotesize{\sffamily{\thepage~\textbar\hspace{3.45cm} 1--\pageref{LastPage}}}}
\fancyhead{}
\renewcommand{\headrulewidth}{0pt} 
\renewcommand{\footrulewidth}{0pt}
\setlength{\arrayrulewidth}{1pt}
\setlength{\columnsep}{6.5mm}
\setlength\bibsep{1pt}

\makeatletter 
\newlength{\figrulesep} 
\setlength{\figrulesep}{0.5\textfloatsep} 

\newcommand{\topfigrule}{\vspace*{-1pt}%
\noindent{\color{cream}\rule[-\figrulesep]{\columnwidth}{1.5pt}} }

\newcommand{\botfigrule}{\vspace*{-2pt}%
\noindent{\color{cream}\rule[\figrulesep]{\columnwidth}{1.5pt}} }

\newcommand{\dblfigrule}{\vspace*{-1pt}%
\noindent{\color{cream}\rule[-\figrulesep]{\textwidth}{1.5pt}} }

\makeatother

\twocolumn[
  \begin{@twocolumnfalse}
 {\includegraphics[height=30pt]{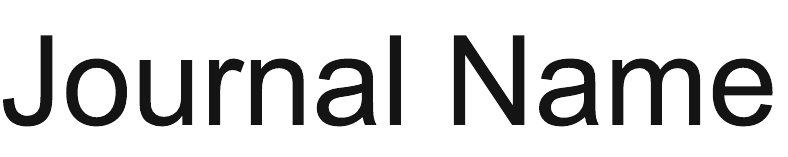}\hfill%
 \raisebox{0pt}[0pt][0pt]{\includegraphics[height=55pt]{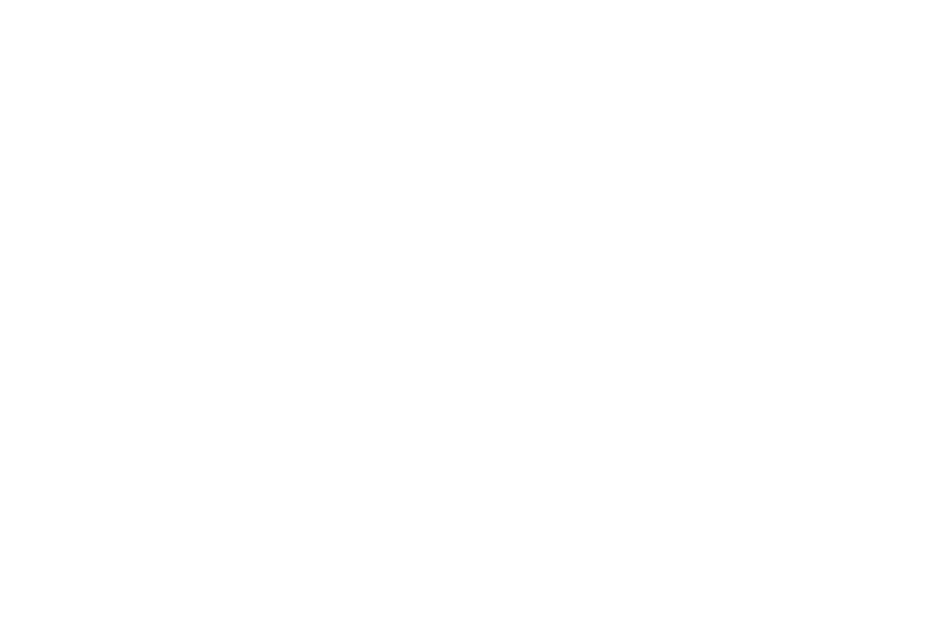}}%
 \\[1ex]%
 \includegraphics[width=18.5cm]{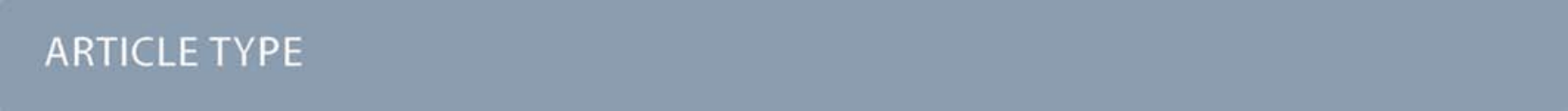}}\par
\vspace{1em}
\sffamily
\begin{tabular}{m{4.5cm} p{13.5cm} }

\includegraphics{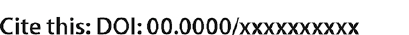} & \noindent\LARGE{\textbf{Impact cratering in sand: Comparing solid and liquid intruders$^\dag$}} \\
\vspace{0.3cm} & \vspace{0.3cm} \\

 & \noindent\large{Rianne de Jong,\textit{$^{a}$} Song-Chuan Zhao,$^{\ast}$\textit{$^{a,b}$}, Diana Garc\'{i}a-Gonz\'{a}lez,\textit{$^{a}$} , Gijs Verduijn,\textit{$^{a}$} and Devaraj van der Meer \textit{$^{a}$}} \\

\includegraphics{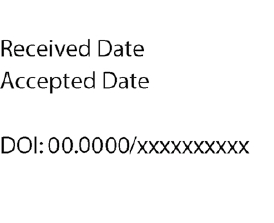} & \noindent\normalsize{How does the impact of a deformable droplet on a granular bed differ from that caused by a solid impactor of similar size and density? Here, we experimentally study this question and focus on the effect of intruder deformability on the crater shape. For comparable impact energies, we show that the crater diameter is larger for droplets than for solid intruders but that the impact of the latter results in deeper craters. Interestingly, for initially dense beds of packing fractions larger than 0.58, we find that the resultant excavated crater volume is independent of the intruder deformability, suggesting an impactor-independent dissipation mechanism within the sand for these dense beds.} \\

\end{tabular}

 \end{@twocolumnfalse} \vspace{0.6cm}

  ]

\renewcommand*\rmdefault{bch}\normalfont\upshape
\rmfamily
\section*{}
\vspace{-1cm}


\footnotetext{\textit{$^{a}$~Physics of Fluids Group, Faculty of Science and Technology, University of Twente, PO Box 217, 7500 AE Enschede, The Netherlands. }}
\footnotetext{\textit{$^{b}$~State Key Laboratory for Strength and Vibration of Mechanical Structures, School of Aerospace, Xian Jiaotong University, Xi'an, Shaanxi, China. }}

\footnotetext{\dag~Electronic Supplementary Information (ESI) available: The ESI contains a comparison between the maximum droplet spreading diameter and the final crater diameter.. See DOI: 00.0000/00000000.}



\newcommand{\etal}{\textit{et al.}}

\newcommand{\ie}{i.e.}
\newcommand{\eg}{e.g.}

\newcommand{\abc}[1]{(\textit{#1})} 
\newcommand{\abt}[1]{\textit{#1}}
\newcommand{\Fig}{Fig.}
\newcommand{\fig}{Fig.}
\newcommand{\figs}{Figs}
\newcommand{\Figs}{Figs}

\newcommand{\zmax}{Z_c^*} 
\newcommand{\Rey}{\mathrm{Re}}
\newcommand{\We}{\mathrm{We}}

\newcommand{\zst}{z_c(r,t^{\infty})} 
\newcommand{\zdyt}{z_c(r,t)} 
\newcommand{\zdy}{z_c(r,t^*)}
\newcommand{\zstxy}{z_c(x,y,t^{\infty})} 
\newcommand{\dcinf}{D_c^{\infty}} 
\newcommand{\dzmax}{D_c^*} 
\newcommand{\Dd}{D_d^*} 

\newcommand{\nonD}{{D_c^{\infty}}/D_0} 
\newcommand{\nonZ}{Z_c^*/D_0} 
\newcommand{\cvol}{V_c^{\infty}} 
\newcommand{\ph}{\phi_0} 

\section{Introduction \label{sd:sec:intro}}
Impacts of objects on sand and soils are regularly observed in nature. This is not only true on the small scale of \eg, raindrops, but equally on the very large scale with asteroid impacts, the craters of which  
are present on the Moon, Mars and even our own planet \cite{Melosh1989, Katsuragi2016}. For planetary craters, it has often been assumed that the intruders are (at least initially) rigid in shape, and accordingly various lab-scale experiments were performed, for which similarities were found between the small craters caused by solid impactors on the laboratory scale and the planetary ones \cite{Holsapple1993, Schmidt1987, Amato1998, Walsh2003}. 
Recently, it was suggested that small scale solid intruders do not quite resemble asteroids, as the latter are expected to fluidize upon hitting 
the target. Pacheco-V\'{a}zquez \& Ruiz-Su\'{a}rez \cite{Pacheco2011} studied impact of consolidated granular intruders on sand that disintegrate upon impact and found similarities with impact craters on the planetary scale \cite{Pacheco2011, Bartali2013}. Even the craters caused by liquid droplet impact were compared to  
large-scale craters \cite{Zhao2015PNAS} and  
it was found that crater dimensions in both cases follow similar scaling laws. 
Although many works indicate the importance of deformability, it remains unknown in what way the consistency of the impactor determines the crater shape. It is precisely this question that we will address in this work, by directly comparing the impact of a solid and a deformable liquid intruder of similar size and density on the laboratory scale.  

A first clear difference between solid and liquid impacts is that 
the droplet continuously deforms upon impact until a maximum droplet spreading diameter is reached \cite{Katsuragi2010, Marston2010, Katsuragi2011, Delon2011, Nefzaoui2012, Emady2013a, Zhao2015PNAS, ZhaoSC2015, Zhao2017, deJong2017}. Consequently, as the droplet deforms upon impact, not all of the impact energy is transferred to the sand bed, but part of it is used for droplet deformation \cite{ZhaoSC2015}. Furthermore, during droplet impact liquid and grains may mix \cite{Katsuragi2010, Delon2011, Emady2013a, ZhaoSC2015}.
We will deal with the latter of these complicating factors by using hydrophobic sand to prevent mixing of liquid and grains, such that we focus solely on the effect of the deformability of the intruder. The first complicating factor is slightly harder to deal with and we will use the approach taken in our earlier work\cite{ZhaoSC2015, Zhao2017, deJong2017} to estimate the portion of the impact energy that is used to excavate the crater from the measured crater depth.  
The craters will be studied using (high-speed) laser profilometry to measure typical crater dimensions such as the maximum crater diameter, the maximum depth and the excavated crater volume.  

\section{Experimental methods and results \label{sd:sec:exp_meth}}
In our experiments we compare two kinds of intruders: the first are deformable millimeter-sized water drops, mixed with food dye to increase contrast (mass fraction $<~2\%$) and the second are solid intruders of similar dimension and density as the water droplets, such that we explore the same range in parameter space for both intruder types. The diameter of the water drops is mostly $2.8$~\si{mm} and occasionally $3.5$~\si{mm}. Three sizes of opaque, spherical solid intruders are used, with diameters of \numlist{2.01; 3.28; 4.00}~\si{mm}, which consist of polypropylene or polyethylene and a white paint coating. The average mass of each intruder size are for increasing diameter, $3.9$~\si{mg}, $17.1$~mg and $36.6$~mg respectively. This results in intruder densities, $\rho_i$, between $900$ and $1100$~\si{kg\per m^3} \cite{footnote1}. 

The impact velocity, $U_0$, is varied in a range of $1.0-5.5$~\si{m/s} by changing the release height and for every droplet is measured just before impact. For the solid intruders the impact velocity is conveniently computed from a calibrated height-velocity profile. 
The target consists of a bed of 
silane-coated (\ie, hydrophobic), soda-lime beads (specific density $\rho_g = \SI{2.5e3}{kg \per m^3}$) of two batches, of which the mean of the size distributions are $d_g = 114$ and $200$~\si{\micro \meter}. 
The packing fraction of the bed, $\ph$, is varied between $0.56$ and $0.62$. Note that the results presented here focus mainly on the $200$~\si{\micro \meter} grains with $\ph=$ \numrange{0.58}{0.60} \cite{footnote2}.
Prior to the experiments, the beads are dried in an oven around a temperature of $\SI{105}{\degreeCelsius}$ for at least half an hour. 
We capture the deformation of the bed surface upon impact with an in-house-built high-speed laser profilometer, from which we acquire a dynamic crater profile, $\zdyt$, by assuming axisymmetry. With the same laser and camera, and the help of a translation stage, both the substrate surface prior to impact and the final crater shape $\zstxy$ are obtained in 3D. By subtracting the initial height profile from the final one, and assuming axisymmetry, we obtain a radial $\zst$ profile, from which we acquire the crater dimensions. We refer to our earlier papers \cite{ZhaoSC2015, deJong2017} for more details on the experimental method. 

\begin{figure}
\centering
	\includegraphics[width=8.6cm]{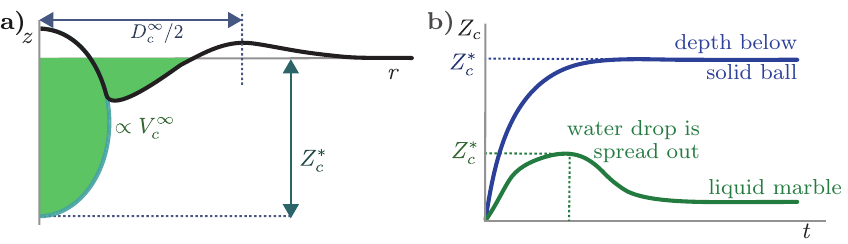}
	\caption{\abc{a} The relevant length scales $\dcinf$, $\zmax$ of the crater and the excavated crater volume $\cvol$ determined from the final crater shape $\zst$. \abc{b} Time evolution of the center crater depth (at $r=0$) for a solid intruder (top line) and water drop  on hydrophobic grains (bottom line). For the latter, the maximum depth is only perceivable during impact, as the droplet contracts after the maximum spreading has reached, causing a subsequent rise of the surface in the center. The maximum depth of a solid intruder will become constant as soon as the object comes to rest.
	\label{sd:fig:defs} }
\end{figure}

\begin{figure*}
\centering
	\includegraphics[width=0.8\textwidth] {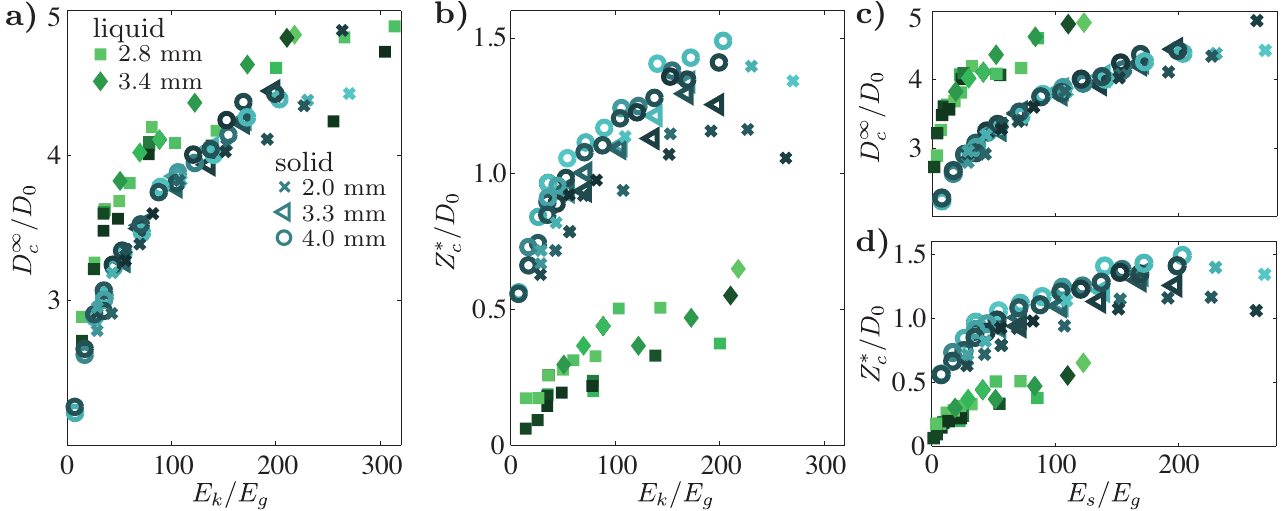}
	\caption{\abc{a},\abc{b} The dimensionless maximum crater diameter $\dcinf/D_0$ \abc{a} and depth $\zmax/D_0$ \abc{b} are plotted against the non-dimensional impact kinetic energy $E_k/E_g$ for deformable water droplets (filled green $\square$, $\lozenge$ symbols) and solid balls (blue $\times$, $\triangleleft$, $\circ$ symbols). The substrate packing fraction is limited to the range $\phi_0 = 0.58-0.60$, where the color of the symbols darkens with increasing $\phi_0$, and the average grain diameter is $d_g= 200$~\si{\micro \meter}.  
\abc{c},\abc{d} The same data sets are plotted as in \abc{a},\abc{b}, but for the droplet data we only take into account the sand deformation energy $E_s$ (\ie, that portion of the impact kinetic energy used for creating the crater; note that $E_s$ is simply equal to $E_k$ for solid impactors). See the main text for precise definitions of $E_s$, $E_k$, and $E_g$. 
	\label{sd:fig:DZ_EkgEsg} }
\end{figure*}

\subsection{Maximum crater diameter. \label{sd:sec:maxdiam}}
The diameter of the final crater shape, $\dcinf$, obtained here as the radial position of the top of the crater rim (see \fig~\ref{sd:fig:defs}\abc{a}), is a typical measure for the target deformation after impact. As the size of solid and liquid intruders varies, we can compare them by non-dimensionalizing the final diameter with the initial intruder diameter $D_0$ and the kinetic energy just upon impact $E_k = (\pi/12) \rho_i D_0^3 U_0^2$ 
with the typical gravitational potential energy $E_g= (\pi/6)\rho_g \phi_0 g D_0^4$, corresponding to a cavity with dimensions $D_0$ dug into the sand \cite{footnote3}.

As the portion of the impact energy that goes into sand deformation is smaller for deformable intruders, intuition suggests that for the same impact speed the resulting crater diameter should be smaller for the droplet than for the solid intruder impacts. In striking contrast to this expectation, one can observe in \fig~\ref{sd:fig:DZ_EkgEsg}\abt{a} that the crater diameter is \emph{larger} when caused by a deformable intruder than by a solid one. This effect becomes even more pronounced when we take into account that, for drop impact, the spreading of the droplet consumes part of the initial impact energy. Using the estimate we derived earlier\cite{ZhaoSC2015, Zhao2017, deJong2017}, the sand deformation energy $E_s$ is given by $E_s = \zmax/(\tfrac{1}{2}D_0 + \zmax) E_k$ for the droplets, where $\zmax$ is the maximum crater depth discussed in greater detail in the next section. With this corrected impact energy (and taking $E_s = E_k$ for the solid intruder data), we find that the difference between crater diameters of solid and liquid impactors becomes even more pronounced (\fig~\ref{sd:fig:DZ_EkgEsg}\abt{c}). So, the simple hypothesis that a crater caused by solid intruders is larger in all directions than the one formed by a droplet does not hold. The opposite is true: the final crater is narrower for solid intruders than for droplets \cite{footnote4}.

How does droplet spreading influence the crater diameter to such an extent that it becomes larger than that of a solid impactor? A first indication could be given by the maximum spreading diameter that the droplet reaches upon impact. In fact, an intuitive approach used in the literature\cite{Katsuragi2010, Delon2011, Katsuragi2016} (and implicitly also in Zhao \etal \cite{Zhao2015PNAS}) relates the crater diameter directly to the maximum droplet spreading diameter. In our experiments we however found that the crater diameter is already developing independently of drop deformation at a quite early stage of the impact \cite{deJong2017}. More specifically, when we compare the final crater diameter $\dcinf$ and maximum droplet spreading diameter $\Dd$ for the same set of experiments, we find that they collapse with different parameters, indicating that the maximum droplet spreading and crater diameters are not directly related to one another, as evidenced in the Supplementary Material\dag 

\begin{figure}\begin{center}
	\includegraphics[width=8.6cm]{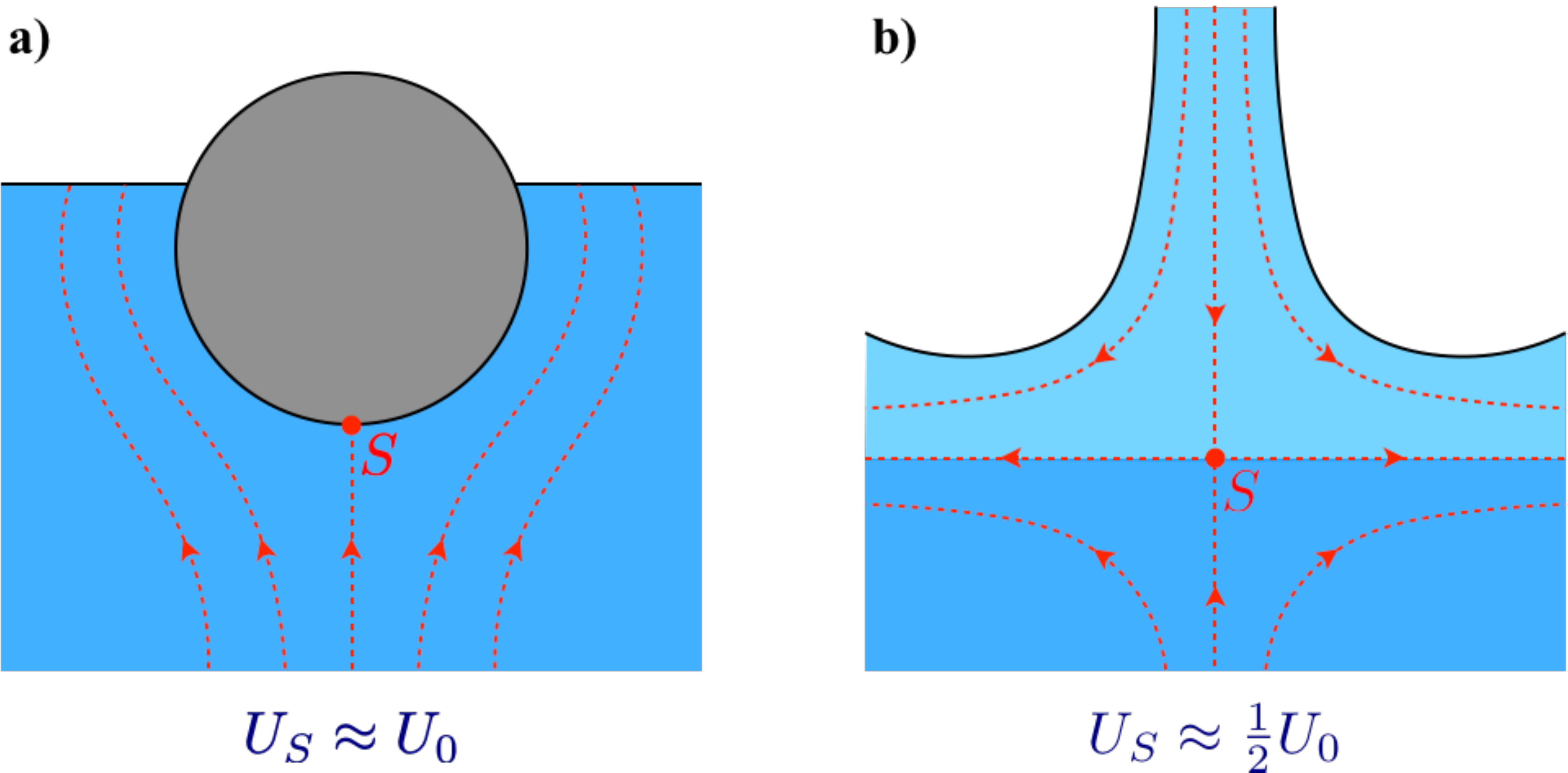}
	\caption{Schematics indicating the difference between solid and liquid impact: A solid impactor creates a dipolar flow in the substrate \abc{a}, whereas a liquid impactor creates a stagnation point flow, similar to that of an impacting jet \abc{b}. Flow patterns are sketched in the frame of reference of the stagnation point, labelled $S$. A key difference arises from the different downward speeds of the stagnation points (equal to the impactor velocity $U_0$ for the solid and $\tfrac{1}{2}U_0$ for the liquid case), which leads to a smaller stagnation pressure for the latter.
	\label{sd:fig:SCHEM} }
\end{center}
\end{figure}

Although droplet spreading cannot directly account for the larger crater diameter for droplet impact, the observed difference between solid and liquid object impact must of course be related to the fact that the droplet will deform under forcing.
It thereby can redirect the downward motion partly into horizontal motion: while the droplet penetrates into the bed, it deforms sideways and pushes grains radially outwards \cite{Pacheco2011}.  
It is instructive to compare the situation with the impact of solid and liquid intruders onto a liquid bath. There, a solid intruder will initially continue with a velocity close to the impact velocity $U_0$ \cite{footnote5},
whereas a droplet will move into the bath with a much smaller initial velocity of maximally $U_0/2$ \cite{footnote6}, 
indicating a rapid transformation of vertical into horizontal motion for the latter. Dynamically, the smaller velocity of the stagnation point in the case of droplet impact corresponds to a smaller stagnation pressure and therefore a smaller vertical forcing.  From a potential flow perspective, as illustrated in \fig~\ref{sd:fig:SCHEM}, for the solid impactor the flow in the liquid bath resembles the dipolar flow around a solid object, whereas for the liquid impactor it resembles a stagnation point flow on the entire scale of the spreading droplet, similar to when a liquid jet impacts a quiescent liquid bath \cite{Bouwhuis2016}. These fundamentally different flow patterns generated in the substrate are (at least transiently) also expected to be present in the case of a sand bed and therefore  may well explain the observed differences in crater diameter.

\subsection{Maximum crater depth \label{sd:sec:maxdepth}}
As the droplet deforms and not all the inertial forcing will be directed downward, we can presume that the penetration depth of a drop will be smaller than that of a solid intruder. For the droplet experiments, the maximum crater depth, $\zmax$, needs to be observed during impact (\fig~\ref{sd:fig:defs}): After the droplet has spread out maximally, the moment at which the crater depth can be measured, it will contract under surface tensional force and become a liquid marble\cite{Aussillous2001}, the presence of which decreases the observed depth of the crater in the center. We therefore obtain $\zmax$ from the dynamic $\zdyt$ profiles. Because for solid intruders the crater depth will remain unaltered as soon as the impactor comes to rest and they rarely get buried completely in our set of experiments, $\zmax$ can simply be found by subtracting the intruder diameter from the central part of the $\zst$ profiles. 

Plotting $\zmax$ as a function of $E_k/E_g$ in~\fig~\ref{sd:fig:DZ_EkgEsg}\abt{b}, we indeed observe 
that for the same impact energy the solid intruder craters become much deeper than 
the ones created by the impact of a droplet. This difference persists when we take into account that  the spreading of the droplet consumes part of the initial impact energy by plotting the same data versus $E_s/E_g$ rather than $E_k/E_g$ (\fig~\ref{sd:fig:DZ_EkgEsg}\abt{d}). 

\begin{figure}\begin{center}
	\includegraphics[width=8.6cm]{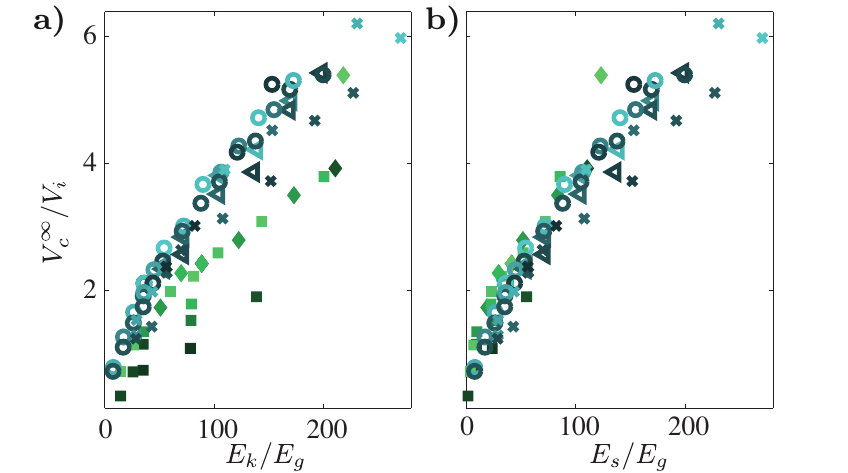}
	\caption{The final crater volume $\cvol$ rescaled with the intruder volume $V_i$, is plotted against \abc{a} the non-dimensional kinetic energy $E_k/E_g$ and \abc{b} the non-dimensionalized portion of the kinetic energy going into sand deformation $E_s/E_g$ for deformable water droplets (filled green $\square$, $\lozenge$ symbols) and solid balls (blue $\times$, $\triangleleft$, $\circ$ symbols). Note that the same data set, colors and symbols as in \fig~\ref{sd:fig:DZ_EkgEsg} are used.
	\label{sd:fig:cVolume} }
\end{center}
\end{figure}

\subsection{Excavated volume \label{sd:sec:final_volume}}
Knowing that exchanging a solid impactor for a deformable one simultaneously increases the crater diameter and decreases its depth makes one wonder what may happen with the total excavated volume. We quantify the crater volume $\cvol$ as the void that is created into the granular bed after impact, \ie, by computing all the granular material that has been displaced from below the original surface of the bed before impact. To do so, we only consider the volume below $z=0$, until the radial distance at which $\zst=0$ is reached (nearby the crater rim), \ie, the colored region sketched in \fig~\ref{sd:fig:defs}\abt{a} \cite{footnote7}.
In \fig~\ref{sd:fig:cVolume}\abt{a}, the excavated crater volume, normalized with the volume of the intruder $V_i = \tfrac{1}{6}\pi D_0^3$, is plotted against the dimensionless impact energy. The crater volume for the solid intruders impacts is 
slightly but noticeably larger for all impact energies than for the droplet intruder cases. If we plot, however, the volume against the corrected impact energy, namely the non-dimensional sand deformation energy $E_s/E_g$, we see that the data nicely collapses onto a single curve. This result shows that despite the fact that the horizontal and vertical crater dimensions differ significantly, the total energy going into crater formation gives rise to the same excavated volume for a solid and a deformable intruder. It suggests that for this set of experiments, the dissipation processes within the sand yield the same result for the dissipated energy during the impact of the two types of intruders, even if the details of the crater formation process are very different. This is even more evident when it is realized that the potential energy stored in the displaced sand volume is found to be less than $5\%$ of the impact energy, \ie, the crater formation process is heavily dominated by dissipation \cite{footnote8}.

\begin{figure}\begin{center}
	\includegraphics[width=8.6cm]{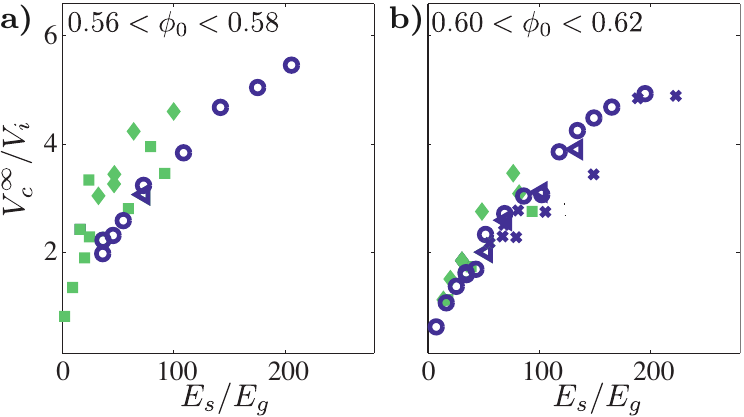}
	\caption{The final crater volume, normalized with the intruder volume $V_i$, is plotted against the non-dimensional kinetic energy $E_s/E_g$  for \abc{a} lower packing fraction, $0.56 \leq \phi_0<0.58$ and \abc{b} higher packing fraction, $0.60\leq\phi_0<0.62$, than the ones reported in \figs~\ref{sd:fig:DZ_EkgEsg} and \ref{sd:fig:cVolume}. The same symbols as in \fig~\ref{sd:fig:DZ_EkgEsg} are used, with the droplet data again filled light green and the solid intruder data with dark purple symbols. 
	\label{sd:fig:cVol_varPhi} }
\end{center}
\end{figure}

\subsection{Packing fraction dependence.\label{sd:subsec:vol_packing}}
Until here, we kept the packing fraction range limited to $0.58 \leq \phi_0 < 0.60$. However, in previous work \cite{deJong2017} on drop impact on sand, we found that the crater shapes and the energy transferred to the sand greatly vary with packing density. Also for solid intruders, the crater shapes are altered with packing fraction as, \eg, discussed by Umbanhowar \& Goldman \cite{Umbanhowar2010}, where it was found that the total displaced volume of sand (including the rim) goes from a compressed state for loose beds (\ie, the total displaced sand volume is smaller than the crater volume), crosses zero at a critical packing fraction ($\phi^* \approx 0.59$), to a dilated state for dense beds. 

So the question is, what happens when we vary the packing fraction? The first observation is that for all attainable packing fraction ranges the crater diameter $\dcinf$ and depth $\zmax$ behave similarly as was found for $0.58 \leq \phi_0 < 0.60$. The results for the excavated volume $\cvol$, however, reveal some subtle differences.
In \fig~\ref{sd:fig:cVol_varPhi}, all excavated volume data $\cvol/V_i$ for packing fraction ranges from $\phi_0 = 0.56 - 0.58$ and $\phi_0 =0.60 - 0.62$ is displayed and we observe that the data for solid and deformable intruders collapses with $E_s/E_g$ for the densest packed beds ($0.60 \leq \ph < 0.62$) as well. For the loose beds with $0.56 \leq \ph <0.58$, however, the droplet data lie slightly but consistently above the crater volumes formed by solid intruders \cite{footnote9}.

What could be the reason for this clear packing fraction dependence? The answer may lie in the way that the sand bed responds, which changes significantly around a critical packing fraction $\phi^*$ \cite{Thompson1991, Schroter2007, Gravish2010, Umbanhowar2010, Metayer2011, Aguilar2016}. Above $\phi^*$ the packing will dilate under forcing \cite{Reynolds1885}, but for loose beds ($\phi_0<\phi^*$) the substrate will respond compressibly until it is (locally) compressed to $\phi^*$, and then start to flow. This is why $\phi^*$ is also known as the dilatancy onset. 

For the loose beds, with $\phi<\phi^*$, this suggests that the substrate will first relatively easily be compressed to $\phi^*$ upon impact, after which the majority of the dissipation between the grains will occur. This may set the difference between the crater volume of the solid and liquid impactor: the droplet can transfer its energy to the bed more effectively as it can move or deform the bed there where it experiences the lowest resistance, and thus the droplet will be a more effective ``compressor", resulting in larger crater volume for the same sand deformation energy $E_s$. Further work is necessary to see if this hypothesis holds, \eg, by impact of slightly deformable solid intruders, (such as hydrogels) of varying stiffness onto beds of varying packing density. 

Finally, we also performed impact experiments on a substrate consisting of slightly smaller hydrophobic, soda-lime beads ($d_g=114$ \si{\micro \meter}), which gave very similar results as reported above. The only difference is that small packing fractions $\phi_0$ present a somewhat larger difference between the dimensionless excavated crater volume $\cvol/V_i$ for droplets and solid impactors (when plotted as a function of $E_s/E_g$) than we reported for $d_g=200$ \si{\micro \meter}. In addition, for the smaller grain size the region in which $\cvol/V_i$ is larger for droplets than for solid impactors extends to a somewhat higher packing fraction.


\section{Conclusions}
We directly compared craters formed by the impact of solid intruders and deformable droplets on a substrate consisting of spherical hydrophobic soda-lime beads. We found that the crater length scales differ for the two types of impactors: solid intruder impact results in deeper but narrower craters compared to the impact of deformable impactors. As explanation we proposed that, whereas for solid intruders the inertial forcing will be directed downward, for droplets the forcing is directed both downwards and sideways. 
This is consistent with the observations of Pacheco-V\'{a}zquez \& Ruiz-Su\'{a}rez \cite{Pacheco2011} for their experiments of consolidated granular sphere impact onto a granular substrate. They observe that, as soon as their granular intruder breaks into pieces, \ie, its yield strength is overcome, the final crater becomes broader and shallower than for solid intruders of the same mass. Where they, however, find that the difference between the crater diameter of the granular intruder and the solid one is constant and independent of impact energy, in our case, the diameters of the solid and the deformable intruder impacts follow different power-law scalings. The exponent $\alpha$ for a fit to the functional form $\nonD \propto (E_s/E_g)^{\alpha}$ is for the water drops $\alpha \approx 0.16$ and for the solid intruders $\alpha \approx 0.21$. In addition, Pacheco-V\'{a}zquez \& Ruiz-Su\'{a}rez \cite{Pacheco2011} found that the depth of the crater formed by a granular object goes towards a relatively constant value as soon as the intruder falls apart, whereas for our drops the crater depth increases with impact energy, see \fig~\ref{sd:fig:DZ_EkgEsg}\abt{b}. 

Subsequently, we studied the resultant excavated crater volume and found surprisingly, that, for dense beds of $\phi_0>0.58$, it is independent of the chosen intruder type. Since these types of impact are heavily dominated by dissipation, this indicates that the amount of dissipation with these dense sand beds can be characterized by the final excavated volume, irrespective of the detailed impact dynamics. Such a one-to-one relation between the excavated volume and the effective impact energy might provide a measure of the energy transfer for a deformable intruder impact by comparing its solid counterpart.

For loosely packed sand beds, we do observe an intruder-type dependence in the crater volume data, which we tentatively attribute to sand compressibility for these low packing fractions. Further research is needed to validate this idea.

\section*{Conflicts of interest}
There are no conflicts to declare.

\section*{Acknowledgements}
We thank Maarten Hannink and Stefan van der Vegte for their help in the solid intruder experiments. This work is financed by the Netherlands Organisation for Scientific Research (NWO) through a VIDI Grant No. 68047512.

\balance


\bibliography{referencesALL.bib} 

\begin{mcitethebibliography}{37}
\bibitem[foo()]{footnote1}
The solid intruders are painted white to make them opaque, such that they cause
  sufficient reflection of the laser light. The spheres of $2.01$ and $4.00$
  ~mm consist of polypropylene and those of $3.28$~mm are made from
  polyethylene. For every size, we painted a batch of spheres, as the intruders
  are small and therefore easily lost during the experiments. The variation in
  density between individual spheres is mostly due to variation in the amount
  of paint on each of them, which is accounted for when calculating the impact
  energy of the intruder.\relax
\mciteBstWouldAddEndPunctfalse
\mciteSetBstMidEndSepPunct{\mcitedefaultmidpunct}
{}{\mcitedefaultseppunct}\relax
\EndOfBibitem
\bibitem[foo()]{footnote2}
Note that the dimensions of the impact crater are expected to depend on $\ph$
  and $d_g$ as the plastic properties of the granular bed (e.g., stiffness and
  friction) depend on those.\relax
\mciteBstWouldAddEndPunctfalse
\mciteSetBstMidEndSepPunct{\mcitedefaultmidpunct}
{}{\mcitedefaultseppunct}\relax
\EndOfBibitem
\bibitem[foo()]{footnote3}
The dimensionless energy $E_k / E_g$ can also be written as a density ratio
  times the Froude number, \ie, $E_k / E_g = \tfrac{1}{2} [\rho_i/(\rho_g
  \phi_0)][U_0^2/(g D_0)]$.\relax
\mciteBstWouldAddEndPunctfalse
\mciteSetBstMidEndSepPunct{\mcitedefaultmidpunct}
{}{\mcitedefaultseppunct}\relax
\EndOfBibitem
\bibitem[foo()]{footnote5}
More specifically, upon impact, the liquid will be set into motion by the
  intruder, giving rise to an added mass force that will decelerate the
  intruder. Assuming a similar density for intruder and target, the velocity of
  the intruder together with the virtual added mass can be approximated at
  $\geq \tfrac{2}{3}U_0$.\relax
\mciteBstWouldAddEndPunctfalse
\mciteSetBstMidEndSepPunct{\mcitedefaultmidpunct}
{}{\mcitedefaultseppunct}\relax
\EndOfBibitem
\bibitem[foo()]{footnote7}
Note that the intruder volume is of course subtracted. For droplet impacts, we
  assume that the liquid volume that may have splashed away is not
  significantly affecting this calculation.\relax
\mciteBstWouldAddEndPunctfalse
\mciteSetBstMidEndSepPunct{\mcitedefaultmidpunct}
{}{\mcitedefaultseppunct}\relax
\EndOfBibitem
\bibitem[foo()]{footnote9}
Remarkably, for these low packing densities, the data seems to collapse with
  the \emph{uncorrected} dimensionless impact energy $E_k/E_g$. We postulate
  that this is a coincidence since the energy consumed by drop deformation
  varies between $30\%$ and $80\%$ of the impact kinetic energy in this set of
  experiments.\relax
\mciteBstWouldAddEndPunctfalse
\mciteSetBstMidEndSepPunct{\mcitedefaultmidpunct}
{}{\mcitedefaultseppunct}\relax
\EndOfBibitem
\end{mcitethebibliography}
\bibliographystyle{rsc} 

\end{document}